\documentclass[a4paper]{article}
\usepackage[T1]{fontenc}
\usepackage[utf8]{inputenc}
\usepackage[italian,english]{babel}
\usepackage{hyperref}
\usepackage[top=1in, bottom=1in, left=1.25in, right=1.25in]{geometry}
\usepackage{natbib}
\usepackage[dvipsnames]{xcolor}
\usepackage{amsmath} 
\usepackage{amssymb} 
\usepackage{amsfonts} 
\usepackage{bm} 
\usepackage{multirow}
\usepackage{graphicx}
\usepackage{pdflscape}
\usepackage{comment}
\usepackage{appendix}
\usepackage{longtable}
\usepackage{enumitem}

\DeclareMathOperator*{\argmax}{arg\,max}

\begin{document}
	
	\title{Bayesian Hierarchical Copula Models with a Dirichlet--Laplace Prior}
	\author{Paolo Onorati \\{MEMOTEF, Sapienza University of Rome} \and Brunero Liseo \\{MEMOTEF, Sapienza University of Rome}}
	\date{}
	\maketitle
	
\begin{abstract}
	We discuss a Bayesian hierarchical copula model for clusters of financial time series. A similar approach has been developed in recent paper.
	However, the prior distributions proposed there do not always provide a proper posterior. In order to circumvent the problem, we adopt a proper global--local shrinkage prior, which is also able to account for potential dependence structures among different clusters. The performance of the proposed model is presented via simulations and a real data analysis.
\end{abstract}

\section{Introduction} \label{sec1}

There is a large body of  literature with respect to hierarchical model settings. The concept to pull the mean of a single group towards the mean across different groups can be found at least in \citet{kelley1927}. \citet{tiao1965} and \citet{hill1965} consider the one-way random effects model and they discuss a Bayesian approach for the analysis of variance because the frequentist unbiased estimator of the variance of random effects could be negative. 
For the same model, \citet{stone1965} discuss and resolve a paradox that arises with the use of  Jeffreys' prior. The foundation for the Bayesian hierarchical linear model is established in \citet{lindley1972}. More recently, \citet{gelman2006} discuss a review on prior distributions for variance parameters in the hierarchical model. 

More recently, \citet{zhuang2020} introduced a hierarchical model in a copula framework; they suggest using, for the variance parameters of two different priors, (i) the standard improper prior for scale parameters, which is proportional to $\sigma^{-2}$, or (ii) a vaguely informative prior, say an inverse gamma density with both parameters equal to a small value.

However, both the above proposals might be impractical: in the first case, the posterior is simply not proper (as we show in the Appendix); in the second case, the use of small parameters of the inverse Gamma priors simply hides the problem without actually solving it; see for example \citet{berger2006}.

\citet{hobert1996} also provide another review on the effect of improper priors in the Gibbs sampling algorithm. 

In this paper, we propose a Bayesian hierarchical copula model using a different prior. In particular, we adopt a global--local shrinkage prior. These prior distributions naturally arise in a linear regression framework with high dimensional data and where a sparsity constraint is necessary  for the vector of coefficients. Several different global--local shrinkage families of priors have been proposed:  \citet{park2008} and \citet{hans2009} discuss the Bayesian LASSO; \citet{carvalho2009} introduce the \textit{Horseshoe prior}, \mbox{\citet{armagan2013}} propose a Generalized Double Pareto prior. 
Here, we will use a Dirichlet--Laplace prior, proposed in \citet{bhattacharya2016}, with a slight modification;
while in a regression framework, it is natural to adopt a prior that shrinks the parameters towards zero, this is not the case for our hierarchical copula model, where the zero value does not have a particular interpretation in the model. For this reason we need to introduce a further level of hierarchy, assuming a prior distribution on the location of the shrinkage point.

The rest of this paper is organized as follows: The next section is devoted to illustrating the statistical model and the prior distribution, highlighting the differences with the approach described in \citet{zhuang2020}; we conclude the section with a description of the sampling algorithm.
In the third section, we perform a simulation study in order to compare the mean square error of the estimates produced by our model and compare them with a standard maximum likelihood approach. Then, we reconsider a dataset discussed in \citet{zhuang2020} and compare the results of the two approaches. We conclude with another illustration of the model in the problem of clustering  financial time series.

\section{Materials and Methods}\label{sec2}
\subsection{The Statistical Model}
\subsubsection{Likelihood and Priors Distributions}
Copula representation is a way to recast a multivariate distribution in such a way that the dependence structure is not influenced by the shape, the parametrization, and the unit of measurement of the marginal distributions.
Their applications in statistical inferences and a review on the most popular approaches can be found in \citet{hofert2015}.
In this paper we will consider several different parametric forms of copula functions: In particular, in the bivariate case, we will use the standard Archimedean families, namely the Joe, Clayton, Gumbel, and Frank copulae.
For more than two dimensions, we will concentrate on the use of the most popular elliptical versions, namely the Gaussian and  Student's $t$ copulae.
Since the main objective of the paper is the clusterization of the dependence structure, for the sake of simplicity and without a loss of generality , we will assume that all  marginal distributions are known or, equivalently, their parameters have been previously estimated. In this way, we can directly work with the transformed variables: 
$U_j = F_{X_j}(x_j)$, $j \in \{1, \dots, n\}$.

Let $c_i(\cdot \vert \psi_i)$ be the generic  copula density function associated with the $i$-th group . The statistical model can be stated as follows:
\begin{equation*}
\big( U_{1i}, U_{2i}, \dots, U_{d_ii} \big) \vert \psi_i \sim c_i(\cdot \vert \psi_i) \ \ \quad \, i \in \{1, \dots, m\}
\end{equation*}
where $m$ denotes the number of groups or clusters. Set  the following:
\begin{equation*}
\gamma_i = \log \bigg(\frac{\psi_i-b_i}{B_i-\psi_i} \bigg) ,
\end{equation*}
and assume the following.
\begin{align*}
\gamma_i \vert \xi, \tau, \alpha_i &\overset{ind}\sim Laplace(\xi, \tau\alpha_i) \ \ \quad \, i \in \{1, \dots, m \}\, , \\
\tau &\sim Gamma \bigg(ma, \frac{1}{2} \bigg) \, , \\
( \alpha_1, \alpha_2, \dots, \alpha_m ) &\sim Dirichlet(a, a, \dots, a) \, , \\
\xi &\sim {Logistic}(0,1).
\end{align*}

In the previous expressions, $b_i$ and $B_i$, respectively, denote the lower and the upper bound of the parameter space of the corresponding $\psi_{i}$, and  $\gamma_{i}$ is the mapping of  $\psi_{i}$ into the real axis; $d_i$ is the dimension of $i$-th group, and $a$ is a hyperparameter, which we typically set to $1$, although different values can be used.
In general, the Archimedean copulae are parametrized in terms of  Kendall's Tau, for which its range of values has been restricted to $(0,1)$ for the Clayton, Joe, and Gumbel copulae, while it is  set to $(-1,1)$ for the Frank copula. 
In the elliptical case, the Gaussian copula is parametrized in terms of the correlation coefficient $\rho$, which ranges in $(-1, 1)$; finally,  Student's $t$ copula has the additional parameter $\nu$, and that is the number of degrees of freedom: A discrete uniform prior on $\{1,2,\dots,35\}$ has been used here.  When  dimension $d$ of the specific group is larger than two, we restrict the analysis to elliptical copulae with an equi-correlation matrix: in that case, it is well known that  the range of the correlation parameter is  $(-1/(d-1), 1)$.

Let $\bm{U}$ be entire observed sample and let $U_{ijk}$ be the $k$-th observation of $i$-th component in the $j$-th group, and let $n_j$ be the number of observation in the $j$-th group. The posterior distribution on the parameter vector $(\bm{\gamma}, \xi, \bm{\alpha}, \tau)$ 
is then described as follows: 
\begin{align*}
\mathrm{p}\big(\bm\gamma, \xi, \bm\alpha, \tau \vert U \big) \propto \prod_{i = 1}^m \bigg[ \prod_{j = 1}^{n_i} \Big[ c_i(U_{1ij}, U_{2ij}, \dots, U_{d_iij} \vert \gamma_i) \Big] &\mathrm{p}(\gamma_i \vert \xi, \tau, \alpha_i) \bigg] \mathrm{p}(\xi) \mathrm{p}(\tau) \mathrm{p}(\bm\alpha),
\end{align*}
where $\bm\gamma = (\gamma_1, \gamma_2, \dots, \gamma_m)$ and $\bm\alpha = (\alpha_1, \alpha_2, \dots, \alpha_m)$. 

The complex form of the posterior distribution requires the use of simulation based methods of inference.
In particular, we will adapt the algorithm of \citet{bhattacharya2016} with a minor modification for the updates of $\bm\gamma$ and the shrinkage location $\xi$.
Following,\citet{bhattacharya2016}, we introduce a vector $\bm\beta =\big(\beta_1, \beta_2, \dots, \beta_m \big) \in \mathbb{R}^m$ in order to have a latent variable representation of the $\bm\gamma$ prior; then, the following is obtained. 
\begin{align*}
\gamma_i \vert \xi, \tau, \alpha_i, \beta_i \overset{ind}\sim& Normal(\xi, \beta_i\tau^2\alpha_i^2) \ \ \forall \, i \in \{1, \dots, m\} \, , \\
\beta_i \overset{iid}\sim& Exp \bigg(\frac{1}{2}\bigg)  \ \ \quad  \, i \in \{1, \dots, m\} .
\end{align*}
\noindent
Here, we briefly describe the algorithm. 
Start the chain at time $0$ by drawing a sample from the prior. At time $t$, we use the following updating procedure:
\begin{enumerate}
	\item Update
 $\bm\gamma \vert \xi, \tau, \bm\alpha, \bm\beta$:
	\begin{enumerate}
		\item Sample $\tilde{\gamma}_i $ from a proposal Cauchy$(\gamma_{it}, \delta_\gamma) \ \ \quad \, i \in \{1, \dots, m\}$;
		\item Set  $\bm{\tilde{\gamma}} = (\tilde{\gamma}_1, \tilde{\gamma}_2,\dots, \tilde{\gamma}_m)$ and
		compute the following.
		\begin{equation*}
		\mathrm{q} = \frac{\prod_{i = 1}^m \bigg[ \prod_{j = 1}^{n_i} \Big[ c_i(U_{1ij}, U_{2ij}, \dots, U_{d_iij} \vert \tilde{\gamma}_i) \Big] \mathrm{p}(\tilde{\gamma}_i \vert \xi_t, \tau_t, \alpha_{it}, \beta_{it}) \bigg]}{\prod_{i = 1}^m \bigg[ \prod_{j = 1}^{n_i} \Big[ c_i(U_{1ij}, U_{2ij}, \dots, U_{d_iij} \vert \gamma_{it}) \Big] \mathrm{p}(\gamma_{it} \vert \xi_t, \tau_t, \alpha_{it}, \beta_{it}) \bigg]}
		\end{equation*}
		\item Sample $u\sim U(0,1)$,
		\item Set $\bm{\gamma_{t+1}} = \bm{\tilde{\gamma}}$ if $u \le q$; otherwise, $\bm{\gamma_{t+1}} = \bm{\gamma_t}$.
	\end{enumerate}
	\item Update $\xi \vert \bm\gamma, \tau, \bm\alpha, \bm\beta$:
	\begin{enumerate}
		\item Sample $\tilde{\xi}$ from a proposal Cauchy$(\xi_t, \delta_{\xi})$;
		\item Compute the following. 
		\begin{equation*}
		q = \frac{\prod_{i=1}^m \Big[ \mathrm{p}(\gamma_{i \, t+1} \vert \tilde{\xi}, \tau_t, \alpha_{it}, \beta_{it}) \Big] \mathrm{p}(\tilde{\xi})}{\prod_{i=1}^m \Big[ \mathrm{p}(\gamma_{i \, t+1} \vert \xi_t, \tau_t, \alpha_{it}, \beta_{it}) \Big] \mathrm{p}(\xi_t)}
		\end{equation*}
		\item  Sample $u\sim U(0,1)$;
		\item Set $\xi_{t+1} = \tilde \xi$ if $u \le q$; otherwise, $\xi_{t+1} = \xi_{t}$.
	\end{enumerate}
	\item Update $\tau \vert \bm\gamma, \xi, \bm\alpha, \bm\beta$: sample $\tau_{t+1} \sim GIG \big(0,1,2 \sum_{i=1}^n \frac{|\gamma_{i \, t+1}-\xi_{t+1}|}{\alpha_{it}} \big)$.
	\item Update $\bm\alpha \vert \bm\gamma, \xi, \tau, \bm\beta$: sample $\tilde \alpha_i \sim GIG(0,1,2|\gamma_{i \, t+1}-\xi_{t+1}|) \ \ \quad \, i \in \{1, \dots, m\}$, and set the following.
	\begin{equation*}
	\alpha_{i \, t+1} = \frac{\tilde \alpha_i}{\sum_{j=1}^m \tilde \alpha_j} \ \ \quad \, i \in \{1, \dots, m\}
	\end{equation*}
	\item Update $\beta_i \vert \bm\gamma, \xi, \tau, \bm\alpha \ \quad \, i  \in \{1, \dots, m\}$: sample $\tilde \beta_i \sim IG(\frac{\tau_{t+1} \alpha_{i \, t+1}}{|\gamma_{i \, t+1}-\xi_{t+1}|}, 1)$ and set  the following.
	\begin{equation*}
	\beta_{i \, t+1} = \frac{1}{\tilde \beta_i} \ \ \quad \  i  \in \{1, \dots, m\} \ .
	\end{equation*}
\end{enumerate}
\noindent
In previous statements, Cauchy$(a,b)$ denotes a one-dimensional Cauchy distribution with location $a$ and scale $b$, while 
$GIG(p,a,b)$ is the generalized inverse Gaussian distribution with the following density function.
\begin{equation*}
f(x) \propto x^{p-1}\exp\left(-\frac{1}{2}ax -\frac{1}{2}\frac{b}{x}\right). 
\end{equation*}
\noindent
Notice that $IG(a,b)$ is the inverse Gaussian distribution, and it is known that  
$X \sim IG(a,b) \Rightarrow X \sim GIG\left(-\frac{1}{2},\frac{b}{a^2}, b \right)$. 
Finally, $\delta_\gamma$ and $\delta_\xi$ are scalar tuning parameters.

In the case of  the Student's t copula,  we need to add another step between stride $1$ and $2$ in order to update $\bm\nu = (\nu_1, \nu_2, \dots, \nu_m)$:
\begin{itemize}
	\item Update $\nu_i \vert \bm\gamma, \xi, \tau, \bm\alpha, \bm\beta \ \ \forall \, i  \in \{1, \dots, m\}$:
	\begin{enumerate}[label = (\alph*)]
		\item Sample $\tilde \nu$ from discrete uniform distribution in $\{1,2, \dots, 35\}$;
		\item Compute the following.
		\begin{equation*}
		\mathrm{q} = \frac{\prod_{j = 1}^{n_i} \Big[ c(U_{1ij}, U_{2ij}, \dots, U_{d_iij} \vert \gamma_{i \, t+1}, \tilde \nu) \Big]}{\prod_{j = 1}^{n_i} \Big[ c(U_{1ij}, U_{2ij}, \dots, U_{d_iij} \vert \gamma_{i \, t+1}, \nu_{it}) \Big]} 
		\end{equation*}
		\item  Sample $u\sim U(0,1)$;
		\item Set $\nu_{i, t+1} = \tilde \nu$ if $u \le q$; otherwise, $\nu_{i, t+1} = \nu_{it}$.
	\end{enumerate}
\end{itemize}

\subsubsection{Prior Distribution of \texorpdfstring{$\boldsymbol{\xi$}}{xi}}
The choice of the prior distribution for the shrinkage location $\xi$ needs some explanation. First of all, notice that, 
according to our prior specification, 
\begin{equation*}
\mathrm{P}(\gamma_i \le \xi) = \frac{1}{2} \ \ \, \quad i  \in \{1, \dots, m\};
\end{equation*}
however $\gamma_i = \log \bigg(\frac{\psi_i-b_i}{B_i-\psi_i} \bigg)$, so otherwise is the case.
\begin{equation*}
\mathrm{P} \bigg(\psi_i \le \frac{B_ie^\xi+b_i}{1+e^\xi} \bigg) = \frac{1}{2}.
\end{equation*}
\noindent

Therefore, given $\xi$, the median of $\psi_i$ is $Y_i= (B_ie^\xi+b_i)/(1+e^\xi) \, \forall i \in \{1, \dots, m\}$. 
Then, it is easy to show that the natural choice of a uniform prior on $Y_i\sim U(b_i, B_i)$ for all $i \in \{1, \dots, m\}$ implies a standard logistic density for $\xi$.

\subsubsection{Previous Work}
Apart form the prior specification, the model described in  previous sections is the one proposed by \citet{zhuang2020}. 
We restrict our discussion to the case where each copula expression has one parameter only.
Their prior can be stated as follows.
\begin{align*}
\gamma_i \vert \mu_i, \sigma^2_i \overset{ind}\sim& N(\mu_i, \sigma^2_i) \ \ \quad \, i  \in \{1, \dots, m\} \ \ , \\
\mu_i \vert \lambda, \delta^2 \overset{iid}{\sim}& N(\lambda, \delta^2) \ \ \quad \, i \in \{1, \dots, m\}\ \ , \\
\sigma^2_i \overset{iid}{\sim}& \pi_{\sigma^2}(\cdot) \ \ \quad \, i  \in \{1, \dots, m\} \ \ , \\
\lambda \sim \pi_{\lambda}(\cdot) \, ,& \, \delta^2 \sim \pi_{\delta^2}(\cdot) \ \ .
\end{align*}

\noindent
There is no  unique choice for the distributions of $(\sigma^2, \lambda, \delta)$, although the authors suggest  using weakly informative priors, for example,  inverse gamma densities with small hyperparameters values or, as an alternative, an objective prior: for example, an improper uniform prior. 
However, one can prove that, in the second case, the posterior distribution cannot be proper no matter what the sample size is. We show this result in  Appendix \ref{app1}. 
When the posterior distribution is improper, the resulting summary statistics are meaningless. In fact, the Markov Chain implied by the MCMC does not have a limiting distribution so the Ergodic theorem does not hold and the posterior is completely useless.
Moreover, even the first solution is not feasible. In fact, when an improper prior produces an improper posterior, using a vague proper prior can typically hide---not solve---the problem. In these cases, in fact, as shown in \citet{berger2006} (p. 398), the use of a vague prior approximating an improper prior typically concentrates the posterior mass on some boundary of the parameter space.

\section{Results}\label{sec3}
\subsection{Simulation Study}
We compare the performance of our approach with the results based on a maximum likelihood approach in a simulation study. We will use a Student's $t$ copula with an equi-correlation matrix and set the number of groups $m$ equal to five. We repeat the procedure $100$ times; at iteration $j$ for the $i$-th group, we sample the true value $\gamma_{ij}^T$ from a standard normal distribution, the degrees of freedom $\nu_{ij}^T$ are sampled from the prior distribution, and the dimensions $d_{ij}$ of the groups are sampled from the uniform discrete distribution in $\{1,2, \dots, 5\}$. Given the parameters and dimensions of the groups, we sample $20$ observations for each group. 
In the maximum likelihood framework, we estimate  the following:
\begin{equation*}
(\hat \gamma_{ij}^{\mathrm{mle}}, \hat \nu_{ij}^{\mathrm{mle}}) = \argmax \prod_{j = 1}^{20} \Big[ c(U_{1ij}, U_{2ij}, \dots, U_{d_iij} \vert \gamma_i, \nu_i \Big] \ \ \quad  \, i  \in \{1, \dots, 5\} \ \ ,
\end{equation*}
and compute the standard errors.
\begin{equation*}
\widehat{SE}^{\, \mathrm{mle}}_{ij} = \left(\gamma_{ij}^T - \hat \gamma_{ij}^{\mathrm{mle}} \right)^2 \ \ \quad \, i  \in \{1, \dots, 5\} \ .
\end{equation*}

\noindent
In a Bayesian framework, we use the posterior mean as a point estimate, obtained from the use of the MCMC algorithm described above. We  ran six independent chains of $2.5 \times 10^5$ scans, discarded the first $5 \times 10^4$  as a burn-in, and finally computed the $\hat \gamma_{ij}^{\mathrm{Bay}}$ via the sample mean of simulation outputs for all $i  \in \{1, \dots, 5\}$. As a tuning parameters, we set $\delta_\gamma = 10^{-3}$ and $\delta_\xi = 10^{-1}$. Then, we compute the following.
\begin{equation*}
\widehat{SE}^{\, \mathrm{Bay}}_{ij} = \left(\gamma_{ij}^T - \hat \gamma_{ij}^{\mathrm{Bay}}\right)^2 \ \ 
\quad \, i  \in \{1, \dots, 5\} \ .
\end{equation*}

\noindent
Comparison  are performed in terms of the corresponding mean square errors.
\begin{equation*}
\widehat{MSE}^{\, \mathrm{mle}}_i = \frac{1}{100} \sum_{j=1}^{100} \widehat{SE}^{\, \mathrm{mle}}_{ij} \ , \qquad		\widehat{MSE}^{\, \mathrm{Bay}}_i \frac{1}{100} \sum_{j=1}^{100} \widehat{SE}^{\, \mathrm{Bay}}_{ij} \ ,
\end{equation*}

Table \ref{table1} reports  values $\widehat{MSE}^{\, \mathrm{mle}}_i$ against $\widehat{MSE}^{\, \mathrm{Bay}}_i$ for all groups based on 100 simulations.
\begin{table}[ht]
	\centering
	\caption{MSE of the proposed Bayesian Hierarchical Model and of the likelihood based one }
	\label{table1}
	\vspace{0.05 in}
	\begin{tabular}{rrrrrr|r}
		\hline
		& 1 & 2 & 3 & 4 & 5 & mean \\ 
		\hline
		Bayes & 0.1449 & 0.1514 & 0.1104 & 0.1106 & 0.1283 & 0.1291 \\ 
		MLE & 0.1861 & 0.1832 & 0.1251 & 0.1477 & 0.1854 & 0.1655 \\ 
		\hline
	\end{tabular}
\end{table}

\subsection{Real Data Applications}
This section is devoted to the implementation of the method in two different applications. The first one is the same as in \citet{zhuang2020} and we include it for comparative purposes; to this end, we quantify the goodness of fit of the model using a predictive approach based on the conditional version of the Widely Applicable Information Criterion, WAIC, in a hierarchical setting, as discussed in \citet{millar2018}. The second one deals with  clustering  financial time series.

\subsubsection{Column Vertebral Data}
We apply our model to the Column Vertebral Data, available at the UCI Machine Learning Repository. It consists of 60 patients with disk hernia, 150 subjects with
spondylolisthesis, and 100 healthy individuals; data are available for the following variables:  angle of
pelvic incidence (PI), angle of pelvic tilt (PT), lumbar lordosis angle (LL),  sacral slope (SS), pelvic radius (PR), and the degree of spondylolisthesis (DS). As in \citet{zhuang2020}, we adopt the generalized skew-t distribution for the marginals, use a maximum likelihood estimator in order to calibrate the parameters and then transform data via the fitted cumulative distribution function. 
Computations were performed using the R package \texttt{sgt} available on CRAN. 
Table \ref{table2} reports the values of fitted parameters for the marginals.
\begin{table}[ht]
	\centering
	\caption{Fitted parameters for each margin distribution}
	\label{table2}
	\vspace{0.05 in}
	\begin{tabular}{c|cccccc}
		\hline
		Group & Feature & $\mu$ & $\sigma$ & $\lambda$ & $p$ & $q$ \\ 
		\hline
		\multirow{5}{*}{Disk Hernia} & PI & 50.2874 & 13.9408 & 0.9992 & 104.9370 & 50.7792 \\ 
		& PT & 17.3686 & 6.9609 & 0.3137 & 1.8070 & 68.7768 \\ 
		& LL & 32.8948 & 11.7179 & 1.0000 & 5.2906 & 364.8091 \\ 
		& SS & 30.4401 & 7.8546 & -0.1599 & 3.5617 & 1.4520 \\ 
		& PR & 116.5142 & 12.9605 & -0.1742 & 5.9304 & 0.4001 \\ 
		& DS & 2.4849 & 5.4948 & -0.1557 & 1.7725 & 358.2803 \\ 
		\hline
		\multirow{5}{*}{Spondylolisthesis} & PI & 71.6191 & 15.0308 & -0.0261 & 1.6375 & 67.3817 \\ 
		& PT & 20.7980 & 11.4766 & 0.2862 & 1.9411 & 44.5023 \\ 
		& LL & 64.0920 & 16.3405 & 0.2633 & 2.1057 & 73.7317 \\ 
		& SS & 49.5130 & 13.1427 & 0.3057 & 46.4772 & 0.0649 \\ 
		& PR & 114.6216 & 15.5666 & 0.0259 & 1.4962 & 32.5924 \\ 
		& DS & 51.6375 & 52.3930 & 0.5757 & 42.0584 & 0.0520 \\ 
		\hline
		\multirow{5}{*}{Healthy} & PI & 51.5086 & 12.4646 & 0.6837 & 2.5388 & 24.2468 \\ 
		& PT & 12.8140 & 6.7551 & -0.1121 & 1.7036 & 71.8428 \\ 
		& LL & 44.9715 & 187.1274 & 0.3583 & 28.3301 & 0.0707 \\ 
		& SS & 38.8785 & 9.6135 & 0.2867 & 1.9040 & 17.9808 \\ 
		& PR & 124.0712 & 53.4395 & 0.1274 & 55.3812 & 0.0364 \\ 
		& DS & 2.1427 & 6.1430 & 0.3069 & 1.2030 & 7.8901 \\ 
		\hline
	\end{tabular}
\end{table}

Following \citet{zhuang2020}, 
we consider the same parametric copulae for the bivariate distributions of the features of interest,  and
for each of these, we construct our Bayesian hierarchical copula model for  three groups of subjects. We run six independent chains of $2.5 \times 10^6$ simulations and discard the first $5 \times 10^5$.  We also set $\delta_\gamma = 10^{-3}$ and $\delta_\xi = 10^{-1}$. 
We did not report any convergence issues, and the multiple Gelman--Rubin test scores for each of the six implemented models \citet{gelman1992} were very close to the optimal value 1.
In terms of the goodness of fit, we have computed the WAIC index for all six models. Our findings is that the most significant relation is the one between PI and PT.
Table \ref{table3} compares the results of \citet{zhuang2020} (model A) with our ones (model B).
The main difference between the results obtained with the two methods is related to the posterior uncertainty quantification. Credible intervals obtained with model B are systemically larger than those obtianed with model A. Our feeling is that it depends on the fact that results in model A are obtained by running a chain where some hyperparameters are fixed to some estimated values, as explained in \citet{zhuang2020}. Fixing values of the hyperparameters eliminates a critical source of variation, inducing  shrinkage in credible intervals size.   

For  the ease of comparisons, we follow \citet{zhuang2020} and report the results not in terms of 
parameter $\gamma$ but rather according the natural parameter of each copula, that is, $\rho$ for the Gaussian copula and $\theta$ for the Archimedean ones. 

\begin{landscape}
	\begin{table}[ht]
		\centering
		\caption{Fitted parameters of copulae}
		\label{table3}
		\begin{tabular}{ccc|ccc|ccc}
			\hline
			\multicolumn{3}{c}{} & \multicolumn{3}{c}{Model A} & \multicolumn{3}{c}{Model B} \\
			Group & Features & Copula & Posterior mean & Posterior s.d. & Posterior CI(95\%) & Posterior mean & Posterior s.d. & Posterior CI(95\%) \\ 
			\hline
			\multirow{6}{*}{Disk Hernia} & PI vs PT & Gaussian & 0.696 & 0.046 & (0.599,0.775) & 0.632 & 0.073 & (0.469, 0.751) \\ 
			 & PI vs SS & Gaussian & 0.726 & 0.040 & (0.633,0.793) & 0.680 & 0.076 & (0.506,0.789) \\ 
			 & DS vs PI & Gaussian & 0.161 & 0.098 & (-0.031,0.339) & 0.229 & 0.126 & (-0.041,0.450) \\
			 & DS vs PT & Frank & -0.511 & 0.577 & (-1.489,0.522) & -0.245 & 0.820 & (-1.858,1.340) \\ 
			 & DS vs LL & Gaussian & 0.244 & 0.103 & (0.031,0.435) & 0.265 & 0.109 & (0.037,0.462) \\ 
			 & DS vs PR & Gaussian & -0.055 & 0.113 & (-0.263,0.175) & -0.075 & 0.126 & (-0.315,0.174) \\ 
			 \hline
			\multirow{6}{*}{Spondylolisthesis} & PI vs PT & Frank & 5.718 & 0.505 & (0.599,0.775) & 5.719 & 0.756 & (4.383,7.138) \\ 
			& PI vs SS & Gumbel & 1.729 & 0.099 & (1.554,1.943) & 1.725 & 0.128 & (1.490,1.984) \\
				& DS vs PI & Frank & 3.427 & 0.431 & (2.552,4.245) & 3.674 & 0.867 & (2.447,4.897) \\ 
			& DS vs PT & Survival Clayton & 0.887 & 0.143 & (0.608,1.174) & 1.036 & 0.193 & (0.679,1.422) \\
			& DS vs LL & Frank & 3.230 & 0.426 & (2.437,4.104) & 3.191 & 0.801 & (2.016,4.370) \\ 
			& DS vs PR & Joe & 1.466 & 0.115 & (1.265,1.698) & 1.421 & 0.154 & (1.121,1.734) \\ 
			\hline
			\multirow{6}{*}{Healthy} & PI vs PT & Gaussian & 0.633 & 0.038 & (0.555,0.699) & 0.621 & 0.057 & (0.496,0.717) \\
			& PI vs SS & Gumbel & 2.574 & 0.178 & (2.239,2.910) & 2.552 & 0.235 & (2.115,3.023) \\ 
			& DS vs PI & Frank & 1.822 & 0.430 & (0.936,2.632) & 1.794 & 1.100 & (0.465,3.139) \\
			& DS vs PT & Gaussian & 0.242 & 0.080 & (0.085,0.401) & 0.210 & 0.102 & (-0.000,0.394) \\ 
			& DS vs LL & Frank & 1.409 & 0.570 & (0.335,2.538) & 1.661 & 0.680 & (0.362,2.970) \\ 
			& DS vs PR & Gaussian & -0.111 & 0.093 & (-0.289,0.065) & -0.076 & 0.123 & (-0.310,0.169) \\ 		
		\end{tabular}
	\end{table}
	\end{landscape}

\subsubsection{Financial Data Application}
Grouping financial time series is important for diversification purposes; a portfolio manager should avoid  investing in instruments with a high degree of positive dependence, and clustering procedures allow  the construction of groups according to some specific risk measure. In this way, financial instruments that belong to the same group will show a certain degree of association; however, the strength of dependence within groups may well be  different in different groups. It is then important to assess the strength of the association for each single cluster, and a method to perform this is  to use a hierarchical structure, such as the one discussed in this paper.

As a risk measure, we consider the so-called tail index, which measures the strength of dependence between two variables when one of them takes extremely low values.  Following \citet{deluca2011}, we construct a dissimilarity measure based on the lower tail coefficient. 
Let $(Y_1,Y_2)$ be a bivariate random vector; the lower tail coefficient $\lambda_L$ of $(Y_1,Y_2)$ is defined as follow:
\begin{equation*}
\lambda_L = \lim_{u\to0^+} \mathrm{P}(F_{Y_1}(Y_1) \le u \vert F_{Y_2}(Y_2) \le u),
\end{equation*}
or, equivalently, 
\begin{equation*}
\lambda_L = \lim_{u \to 0^+} \frac{C(u,u)}{u} \ \ ,
\end{equation*}
where $C(\cdot, \cdot)$ is the cumulative distribution function of the copula associated to $(Y_1, Y_2)$.
In order to estimate $\lambda_L$, we use the empirical estimator discussed in  \cite{joe1992}:
\begin{equation*}
\hat{\lambda}_L = \frac{\hat{C}(\frac{\sqrt n}{n},\frac{\sqrt n}{n})}{\frac{\sqrt n}{n}} \ \ ,
\end{equation*}
where $\hat C (\cdot, \cdot)$ is the empirical copula, and $n$ is the sample size.
The dissimilarity measure is then defined as  follows.
\begin{equation*}
d(Y_1,Y_2) = 1-\lambda_L(Y_1,Y_2) \ \ ,
\end{equation*}

\noindent
The preliminary clustering procedure has been implemented using a complete linkage method.
Notice that a bivariate lower tail coefficient is not the unique method for modeling dependence on extreme low values: \citet{durante2014} proposed a conditioned correlation coefficient estimated using a nonparametric approach; \citet{fuchs2021} analyzed dissimilarity measure applicable to a multivariate lower tail coefficient.

We consider the ``S\&P 500 Full Dataset'' available at Kaggle: It contains  more relevant information for the components of 
S\&P 500. We take the daily closing prices from  5 June 2000 to 5 June 2020 and discard instruments without a complete record for this period. Then, we restrict our analysis to 379 components. For all of them, we computed the log-returns by taking  log-differences and filter data by fitting; for each time series,  an ARMA(1,1)GJR-GARCH(1,1) model with Student's $t$ innovations was used; then, we extracted residuals and transformed them via the fitted cumulative distribution function in order to obtain  pseudo-data. Computations were performed using the CRAN package \texttt{rugarch}.  Hence, we compute the empirical estimator of the lower tail coefficient for any possible pair and the dissimilarity measure associated and use them to feed the clustering algorithm. Due to computational complexities, we  used the coarsest partition under the constraint that the largest group must have at most  10 components. We obtained 30 groups with dimensions of more than one and discarded instruments that belong to groups with only one component.
The final number of instruments was thus reduced to 93.

We ran the MCMC algorithm described above for the 30 clusters, performing 12 independent chains of $10^5$ scans and discarding the first $1.5 \times 10^4$ as they burned in. Tuning parameters were set to $\delta_\gamma = 10^{-6}$, $\delta_{\xi} = 10^{-3}$. 
Moreover, in this example, we did not report any convergence issues, and the Gelman--Rubin test score was 1.02.  For each scan and for any group, we compute the lower tail coefficient via the following formula:
\begin{equation*}
\lambda_L = 2 T_{\nu+1} \left( - \sqrt \frac{(\nu+1)(1-\rho)}{1+\rho} \right) \ \ ,
\end{equation*}
where $T_\nu(\cdot)$ is the univariate cumulative distribution function of a Student's $t$ random variable with $\nu$ degrees of freedom. The copula used in this example was a Student's $t$ copula with an equi-correlation matrix: As a consequence, we obtained a single value for the lower tail coefficient for each cluster. Table \ref{table4} reports the results 
for each pair that belongs to the same group. Finally, we report the estimation results.

\begin{center}
	\begin{longtable}{rr|rrr}
		\caption{Posterior distributions for lower tail coefficients}
		\label{table4} \\
		\hline
		Group & Components & Posterior mean & Posterior s.d. & Posterior CI(95\%) \\ 
		\hline
		\endfirsthead
		\hline
		Group & Components & Posterior mean & Posterior s.d. & Posterior CI(95\%) \\ 
		\hline
		\endhead
		\multirow{2}{*}{1} & NTRS & \multirow{2}{*}{0.5001} & \multirow{2}{*}{0.0592} & \multirow{2}{*}{(0.4153, 0.5918)} \\ 
		& STT & & & \\
		\hline
        \multirow{2}{*}{2} & CVX & \multirow{2}{*}{0.4833} & \multirow{2}{*}{0.0592} & \multirow{2}{*}{(0.4061, 0.5715)} \\ 
		& XOM & & & \\
		\hline
		\multirow{2}{*}{3} & AMAT & \multirow{2}{*}{0.4499} & \multirow{2}{*}{0.0633} & \multirow{2}{*}{(0.3648, 0.5573)} \\ 
		& LRCX & & & \\
		\hline
		\multirow{2}{*}{4} & BEN & \multirow{2}{*}{0.4259} & \multirow{2}{*}{0.0649} & \multirow{2}{*}{(0.3457, 0.5359)} \\ 
		& TROW & & & \\
		\hline
        \multirow{2}{*}{5} & CMS & \multirow{2}{*}{0.4256} & \multirow{2}{*}{0.0661} & \multirow{2}{*}{(0.3347, 0.5296)} \\ 
		& PNW & & & \\
		\hline
        \multirow{2}{*}{6} & APD & \multirow{2}{*}{0.4198} & \multirow{2}{*}{0.0655} & \multirow{2}{*}{(0.3389, 0.5274)} \\ 
		& LIN & & & \\
		\hline
		\multirow{3}{*}{7} & PEAK & \multirow{3}{*}{0.4170} & \multirow{3}{*}{0.0636} & \multirow{3}{*}{(0.3538, 0.5097)} \\
		& VTR & & & \\
		& WELL & & & \\
		\hline
		\multirow{3}{*}{8} & DHI & \multirow{3}{*}{0.3942} & \multirow{3}{*}{0.0643} & \multirow{3}{*}{(0.3137, 0.4895)} \\ 
		& LEN & & & \\
		& PHM & & & \\
		\hline
		\multirow{2}{*}{9} & MLM & \multirow{2}{*}{0.3827} & \multirow{2}{*}{0.0678} & \multirow{2}{*}{(0.2881, 0.4963)} \\ 
		& VMC & & & \\
		\hline
		\multirow{2}{*}{10} & HD & \multirow{2}{*}{0.3757} & \multirow{2}{*}{0.0675} & \multirow{2}{*}{(0.2828, 0.4851)} \\ 
		& LOW & & & \\
		\hline
		\multirow{2}{*}{11} & COP & \multirow{2}{*}{0.3685} & \multirow{2}{*}{0.0681} & \multirow{2}{*}{(0.2765, 0.4880)} \\
		& MRO & & & \\
		\hline 
		\multirow{2}{*}{12} & ADP & \multirow{2}{*}{0.3532} & \multirow{2}{*}{0.0692} & \multirow{2}{*}{(0.2663, 0.4704)} \\
		& PAYX & & & \\
		\hline 
		\multirow{3}{*}{13} & CSX & \multirow{3}{*}{0.3395} & \multirow{3}{*}{0.0674} & \multirow{3}{*}{(0.2672, 0.4535)} \\ 
		& NSC & & & \\
		& UNP & & & \\
		\hline
		\multirow{2}{*}{14} & T & \multirow{2}{*}{0.3338} & \multirow{2}{*}{0.0699} & \multirow{2}{*}{(0.2368, 0.4509)} \\ 
		& VZ & & & \\
		\hline
		\multirow{2}{*}{15} & CAH & \multirow{2}{*}{0.3337} & \multirow{2}{*}{0.0691} & \multirow{2}{*}{(0.2414, 0.4401)} \\
		& MCK & & & \\
		\hline
		\multirow{4}{*}{16} & BAC & \multirow{4}{*}{0.3235} & \multirow{4}{*}{0.0671} & \multirow{4}{*}{(0.2590, 0.4203)} \\ 
		& C & & & \\
		& JMP & & & \\
		& MS & & & \\
		\hline
		\multirow{5}{*}{17} & AIV & \multirow{5}{*}{0.3221} & \multirow{5}{*}{0.0668} & \multirow{5}{*}{(0.2593, 0.4187)} \\ 
		& AVB & & & \\
		& EQR & & & \\
		& ESS & & & \\
		& UDR & & & \\
		\hline 
		\multirow{2}{*}{18} & RSG & \multirow{2}{*}{0.3168} & \multirow{2}{*}{0.0694} & \multirow{2}{*}{(0.2275, 0.4255)} \\ 
		& WM & & & \\
		\hline
		\multirow{3}{*}{19} & DVN & \multirow{3}{*}{0.2979} & \multirow{3}{*}{0.0682} & \multirow{3}{*}{(0.2166, 0.4103)} \\ 
		& EOG & & & \\
		& NBL & & & \\
		\hline
		\multirow{2}{*}{20} & D & \multirow{2}{*}{0.2932} & \multirow{2}{*}{0.0708} & \multirow{2}{*}{(0.1953, 0.4113)} \\ 
		& SO & & & \\
		\hline
		\multirow{2}{*}{21} & NI & \multirow{2}{*}{0.2920} & \multirow{2}{*}{0.0700} & \multirow{2}{*}{(0.2022, 0.4032)} \\ 
		& SRE & & & \\
		\hline
		\multirow{2}{*}{22} & IP & \multirow{2}{*}{0.2914} & \multirow{2}{*}{0.0713} & \multirow{2}{*}{(0.1957, 0.4145)} \\ 
		& PKG & & & \\
		\hline
		\multirow{2}{*}{23} & CB & \multirow{2}{*}{0.2839} & \multirow{2}{*}{0.0715} & \multirow{2}{*}{(0.1815, 0.4132)} \\
		& TRV & & & \\
		\hline 
		\multirow{4}{*}{24} & GL & \multirow{4}{*}{0.2818} & \multirow{4}{*}{0.0677} & \multirow{4}{*}{(0.2177, 0.3804)} \\ 
		& LNC & & & \\
		& MET & & & \\
		& UNM & & & \\
		\hline
		\multirow{9}{*}{25} & CMA & \multirow{9}{*}{0.2294} & \multirow{9}{*}{0.0666} & \multirow{9}{*}{(0.1526, 0.3273)} \\ 
		& FITB & & & \\
		& HBAN & & & \\
		& KEY & & & \\
		& MTB & & & \\
		& PNC & & & \\
		& RF & & & \\
		& TFC & & & \\
		& USB& & & \\
		\hline
		\multirow{2}{*}{26} & ATO & \multirow{2}{*}{0.2201} & \multirow{2}{*}{0.0692} & \multirow{2}{*}{(0.1256, 0.3412)} \\
		& EVRG & & & \\
	    \hline
		\multirow{3}{*}{27} & ETR & \multirow{3}{*}{0.1923} & \multirow{3}{*}{0.0652} & \multirow{3}{*}{(0.1175, 0.2953)} \\ 
		& NEE & & & \\
		& PEG & & & \\
		\hline
		\multirow{9}{*}{28} & AEE & \multirow{9}{*}{0.1768} & \multirow{9}{*}{0.0633} & \multirow{9}{*}{(0.1174, 0.2855)} \\ 
		& AEP & & & \\
		& DTE & & & \\
		& DUK & & & \\
		& ED & & & \\
		& ES & & & \\
		& LNT & & & \\
		& WEC & & & \\
		& XEL & & & \\
		\hline
		\multirow{9}{*}{29} & ARE & \multirow{9}{*}{0.1522} & \multirow{9}{*}{0.0605} & \multirow{9}{*}{(0.0874, 0.2439)} \\ 
		& BXP & & & \\
		& DRE & & & \\
		& FRT & & & \\
		& KIM & & & \\
		& MAA & & & \\
		& PLD & & & \\
		& REG & & & \\
		& SPG & & & \\
		\hline 
		\multirow{2}{*}{30} & EW & \multirow{2}{*}{0.0008} & \multirow{2}{*}{0.0011} & \multirow{2}{*}{(0.0000, 0.0028)} \\ 
		& SYK & & & \\
		\hline
	\end{longtable}
\end{center}

\section{Conclusions}\label{sec4}
We discussed and improved a fully Bayesian analysis for a hierarchical copula model proposed in \citet{zhuang2020}. 
We proposed the use of a proper prior, which is able to induce shrinkage and, at the same time, dependence  among different clusters of observations. This prior does not mimic the behavior of an improper prior and is better suited for objectively representing  information coming from the data. 
Our prior belongs to the large family of globa--local shrinkage densities, with an extra stage in the hierarchy, due to the absence of a significant shrinkage value;  we  experienced that this approach is very effective and useful in the case of parametric copulae depending on a single parameter. In a  more general situation, this approach needs to be modified, and this can be easily accommodated. 

Finally, we presented an application in a financial context, where the goal was to estimate the lower tail coefficient of several financial time series in a parametric way using  the Student's $t$ copula.

\nocite{portnoy1971}
\nocite{dua2019}
\nocite{roll2020}
\nocite{sgtPKG_R}
\nocite{rugarchPKG_R}
	
\bibliography{DLESC2.bib}	

\begin{thebibliography}{26}
\providecommand{\natexlab}[1]{#1}
\providecommand{\url}[1]{\texttt{#1}}
\expandafter\ifx\csname urlstyle\endcsname\relax
  \providecommand{\doi}[1]{doi: #1}\else
  \providecommand{\doi}{doi: \begingroup \urlstyle{rm}\Url}\fi

\bibitem[Armagan et~al.(2013)Armagan, Dunson, and Lee]{armagan2013}
Artin Armagan, David Dunson, and Jaeyong Lee.
\newblock Generalized double pareto shrinkage.
\newblock \emph{Statistica Sinica}, 23\penalty0 (1):\penalty0 119--143, 2013.

\bibitem[Berger(2006)]{berger2006}
James Berger.
\newblock The case for objective bayesian analysis.
\newblock \emph{Bayesian Analysis}, 1\penalty0 (3):\penalty0 385 -- 402, 2006.

\bibitem[Bhattacharya et~al.(2016)Bhattacharya, Pati, Pillai, and
  Dunson]{bhattacharya2016}
Anirban Bhattacharya, Debdeep Pati, Natesh~S. Pillai, and David~B. Dunson.
\newblock Dirichlet–laplace priors for optimal shrinkage.
\newblock \emph{Journal of the American Statistical Association}, 110\penalty0
  (512):\penalty0 1479 -- 1490, 2016.

\bibitem[Carvalho et~al.(2010)Carvalho, Polson, and Scott]{carvalho2009}
Carlos~M. Carvalho, Nicholas~G. Polson, and James~G. Scott.
\newblock {The horseshoe estimator for sparse signals}.
\newblock \emph{Biometrika}, 97\penalty0 (2):\penalty0 465--480, 04 2010.

\bibitem[Davis(2015)]{sgtPKG_R}
Carter Davis.
\newblock \emph{\texttt{sgt}: Skewed Generalized T Distribution Tree.}, 2015.
\newblock R package version 2.0.

\bibitem[De~Luca and Zuccolotto(2011)]{deluca2011}
Giovanni De~Luca and Paola Zuccolotto.
\newblock A tail dependence-based dissimilarity measure for financial time
  series clustering.
\newblock \emph{Advances in Data Analysis and Classification}, 5\penalty0
  (4):\penalty0 323--340, 2011.

\bibitem[Dua(2017)]{dua2019}
Graff Dua.
\newblock Uci machine learning repository, 2017.

\bibitem[Durante et~al.(2014)Durante, Pappadà, and Torelli]{durante2014}
Fabrizio Durante, Roberta Pappadà, and Nicola Torelli.
\newblock Clustering of financial time series in risky scenarios.
\newblock \emph{Advances in Data Analysis and Classification}, 8\penalty0
  (4):\penalty0 359--376, 2014.

\bibitem[Fuchs et~al.(2021)Fuchs, {Di Lascio}, and Durante]{fuchs2021}
Sebastian Fuchs, F.~Marta~L. {Di Lascio}, and Fabrizio Durante.
\newblock Dissimilarity functions for rank-invariant hierarchical clustering of
  continuous variables.
\newblock \emph{Computational Statistics {\&} Data Analysis}, 159:\penalty0
  107201, 2021.

\bibitem[Gelman(2006)]{gelman2006}
Andrew Gelman.
\newblock Prior distributions for variance parameters in hierarchical models
  (comment on article by browne and draper.
\newblock \emph{Bayesian Analysis}, 1\penalty0 (3):\penalty0 515--534, 2006.

\bibitem[Gelman and Rubin(1992)]{gelman1992}
Andrew Gelman and Donald~B Rubin.
\newblock Inference from iterative simulation using multiple sequences.
\newblock \emph{Statistical science}, 7\penalty0 (4):\penalty0 457--472, 1992.

\bibitem[Ghalanos(2020)]{rugarchPKG_R}
Alexios Ghalanos.
\newblock \emph{\texttt{rugarch}: Univariate GARCH models.}, 2020.
\newblock R package version 1.4-4.

\bibitem[Hans(2009)]{hans2009}
Chris Hans.
\newblock Bayesian lasso regression.
\newblock \emph{Biometrika}, 96\penalty0 (4):\penalty0 835--845, 2009.

\bibitem[Hill(1965)]{hill1965}
Bruce~M. Hill.
\newblock Inference about variance components in the one-way model.
\newblock \emph{Journal of the American Statistical Association}, 60\penalty0
  (311):\penalty0 806--825, 1965.

\bibitem[Hobert and Casella(1996)]{hobert1996}
James~P. Hobert and George Casella.
\newblock The effect of improper priors on gibbs sampling in hierarchical
  linear mixed models.
\newblock \emph{Journal of the American Statistical Association}, 91\penalty0
  (436):\penalty0 1461--1473, 1996.

\bibitem[Hofert et~al.(2018)Hofert, Kojadinovic, Maechler, and Yan]{hofert2015}
Marius Hofert, Ivan Kojadinovic, Martin Maechler, and Jun Yan.
\newblock \emph{{E}lements of {C}opula {M}odeling with \textsf{R}}.
\newblock Springer Use R! Series, 2018.
\newblock ISBN 978-3-319-89635-9.

\bibitem[Joe et~al.(1992)Joe, Smith, and Weissman]{joe1992}
Harry Joe, Richard~L. Smith, and Ishay Weissman.
\newblock Bivariate threshold methods for extremes.
\newblock \emph{Journal of the Royal Statistical Society. Series B
  (Methodological)}, 54\penalty0 (1):\penalty0 171--183, 1992.

\bibitem[Kelley(1927)]{kelley1927}
Lee~Truman Kelley.
\newblock \emph{The Interpretation of Educational Measurement}.
\newblock Measurement and adjustment series. WorldBook Company,
  Yonkers-on-Hudson, N.Y., 1927.

\bibitem[Lindley and Smith(1972)]{lindley1972}
D.~V. Lindley and A.~F.~M. Smith.
\newblock Bayes estimates for the linear model.
\newblock \emph{Journal of the Royal Statistical Society. Series B
  (Methodological)}, 34\penalty0 (1):\penalty0 1--41, 1972.

\bibitem[Millar(2018)]{millar2018}
Russell~B Millar.
\newblock Conditional vs marginal estimation of the predictive loss of
  hierarchical models using {WAIC} and cross-validation.
\newblock \emph{Statistics and Computing}, 28\penalty0 (2):\penalty0 375--385,
  2018.

\bibitem[Park and Casella(2008)]{park2008}
Trevor Park and George Casella.
\newblock The bayesian lasso.
\newblock \emph{Journal of the American Statistical Association}, 103\penalty0
  (482):\penalty0 681--686, 2008.

\bibitem[Portnoy(1971)]{portnoy1971}
Stephen Portnoy.
\newblock Formal bayes estimation with application to a random effects model.
\newblock \emph{The Annals of Mathematical Statistics}, 42\penalty0
  (4):\penalty0 1379--1402, 1971.

\bibitem[Roll(2020)]{roll2020}
Roll.
\newblock S\&p 500 full dataset, 2020.

\bibitem[Stone and Springer(1965)]{stone1965}
M.~Stone and B.~G.~F. Springer.
\newblock A paradox involving quasi prior distributions.
\newblock \emph{Biometrika}, 52\penalty0 (3/4):\penalty0 623--627, 1965.

\bibitem[Tiao and Tan(1965)]{tiao1965}
George~C. Tiao and W.~Y. Tan.
\newblock Bayesian analysis of random-effect models in the analysis of
  variance. i. posterior distribution of variance-components.
\newblock \emph{Biometrika}, 52\penalty0 (1/2):\penalty0 37--53, 1965.

\bibitem[Zhuang et~al.(2020)Zhuang, Diao, and Yi]{zhuang2020}
Haoxin Zhuang, Liqun Diao, and Grace~Y. Yi.
\newblock A bayesian hierarchical copula model.
\newblock \emph{Electronic Journal of Statistics}, 14(2)\penalty0 (2):\penalty0
  4457 -- 4488, 2020.

\end{thebibliography}

\begin{appendices}
	\section[A]{Appendix} \label{app1}

	Here, we show that the prior proposed in \citet{zhuang2020} leads to an improper posterior

	The statistical model consists of  $m$ $d$-dimensional copulae governing different sets of observations.
\begin{equation*}
\big( U_{1i}, U_{2i}, \dots, U_{d_ii} \big) \vert \theta_i \sim c_i(\cdot \vert \theta_i) \ \ \quad \, i  \in \{1, \dots, m\} \, .
\end{equation*}	

\noindent
Let $\gamma_i = \eta_i g_i(\theta_i)$; here, $\eta_i$ is a scaling parameter that can be considered known.  One-to-one mapping functions  $g_i(\cdot)$ are needed to put all  dependence parameters on the real line.
\citet{zhuang2020} made the following assumptions.
\begin{align*}
\gamma_i \vert \mu_i, \sigma^2_i \overset{ind}{\sim}& N(\mu_i, \sigma^2_i) \ \ \quad \, i  \in \{1, \dots, m\}; \\
\mu_i \vert \lambda, \delta^2 \overset{iid}{\sim}& N(\lambda, \delta^2) \ \ \quad \, i  \in \{1, \dots, m\}.
\end{align*}

\noindent
Hyper-parameters $\sigma_i$'s, $\lambda$, and $\delta^2$ are given a suitable prior distribution. 
For the moment, we do not specify the priors and set the following.
\begin{align*}
\sigma^2_i \overset{iid}{\sim} \pi_{\sigma^2} (\cdot)& \ \ \quad \, i \in \{1, \dots, m\}. \\
\lambda \sim \pi_{\lambda}(\cdot) \, , & \,
\delta^2 \sim \pi_{\delta^2}(\cdot)  \ \ .
\end{align*}

\noindent
Since the $g_i(\theta_i)$'s are one-to-one,  we write $c_i(\cdot \vert \gamma_i)$ instead of $c_i(\cdot \vert \theta_i)$. 
Let $U$ be the observed sample, and let $U_{ijk}$ be the $k$-th observation of $i$-th component in the $j$-th group. Let $n_j$ be the sample size of the $j$-th group.
Furthermore, 
let  $\bm{\gamma} = (\gamma_1, \gamma_2, \dots, \gamma_m), \bm{\mu} = (\mu_1, \mu_2, \dots, \mu_m)$, and $\bm{\sigma^2} = (\sigma^2_1, \sigma^2_2, \dots, \sigma^2_m)$.
Finally, let $S(\omega)$ denote the parameter space of the generic parameter $\omega$.

The next proposition shows that, using standard noninformative priors for scale and location parameters, 
the resulting posterior will be improper independently of the sample size.

{\bf Proposition 1:} If $\pi_{\sigma^2_i}(\sigma_i^2) \propto \sigma^{-2}_i,$ for $i \in \{1, \dots, m\}$, and $\pi_{\delta^2}(\delta^2) \propto\delta^{-2}, \ \, \pi_{\lambda}(\lambda) \propto 1$, the posterior distribution $\bm{\gamma} \vert U$ is  improper for any choice of the copula densities   $c_i(\cdot \vert \gamma_i )$ and independently of the sample size.

\noindent
{\bf Proof:}
For the sake of clarity, set $d\bm{\sigma}^2 = d\sigma_1^2 d\sigma_2^2 \dots d\sigma^2_m$ and 
$d\bm{\mu} = d\mu_1 d\mu_2 \dots d\mu_m$.
We need to show that the following pseudo-marginal posterior distribution of $\bm{\gamma}$ is not integrable:
\begin{align*}
\pi(\bm\gamma \vert U) =& \int_{S(\bm\mu)} \int_{s(\bm{\sigma^2})} \int_{S(\delta^2)} \int_{S(\lambda)} \pi(\bm\gamma, \bm\mu, \bm{\sigma^2}, \lambda, \delta^2 \vert U) d\lambda d\delta^2 d\bm{\sigma}^2 d\bm\mu \\
\propto& \int_{S(\bm\mu)} \int_{s(\bm{\sigma^2})} \int_{S(\delta^2)} \int_{S(\lambda)} \pi(U \vert \bm\gamma, \bm\mu, \bm{\sigma^2}, \lambda, \delta^2) \pi(\bm\gamma, \bm\mu, \bm{\sigma^2}, \lambda, \delta^2) d\lambda d\delta^2 d\bm{\sigma}^2 d\bm\mu\ \ ,
\end{align*}
where $\pi(U \vert \bm\gamma, \bm\mu, \bm{\sigma^2}, \lambda, \delta^2)$ represents the likelihood function.  Then, we obtain the following:
\begin{align*}
\pi(\bm\gamma \vert U) \propto& \int_{S(\bm\mu)} \int_{s(\bm{\sigma^2})} \int_{S(\delta^2)} \int_{S(\lambda)} \prod_{i=1}^{m} \Big[ \prod_{j=1}^{n_i} \Big( c_i(U_{1ij}, U_{2ij}, \dots U_{d_iij} \vert \gamma_i)  \Big) \Big] \times \\
& \pi(\bm\gamma \vert \bm\mu, \bm{\sigma^2}) \pi(\bm\mu \vert \lambda, \delta^2) \pi(\bm{\sigma^2}) \pi(\lambda) \pi(\delta^2) d\lambda d\delta^2 d\bm{\sigma}^2 d\bm\mu \\
\propto& \prod_{i=1}^{m} \Big[ \prod_{j=1}^{n_i} \Big( c_i(U_{1ij}, U_{2ij}, \dots U_{d_iij} \vert \gamma_i)  \Big) \Big] \int_{S(\bm\mu)} \int_{s(\bm{\sigma^2})} \pi(\bm\gamma \vert \bm\mu, \bm{\sigma^2}) \pi(\bm{\sigma^2}) \times \\
&\int_{S(\delta^2)} \int_{S(\lambda)} \pi(\bm\mu \vert \lambda, \delta^2) \pi(\lambda) \pi(\delta^2) d\lambda d\delta^2 d\bm{\sigma}^2 d\bm\mu \\
=& \prod_{i=1}^{m} \Big[ \prod_{j=1}^{n_i} \Big( c_i(U_{1ij}, U_{2ij}, \dots U_{d_iij} \vert \gamma_i) \Big) \Big] \pi(\bm\gamma) \ \ ,
\end{align*}
with 
\begin{align*}
\pi(\bm\gamma) =& \int_{S(\bm\mu)} \int_{s(\bm{\sigma^2})} \pi(\bm\gamma \vert \bm\mu, \bm{\sigma^2}) \pi(\bm{\sigma^2}) \pi(\bm\mu) d\bm{\sigma}^2 d\bm\mu 
\end{align*}
and
\begin{align*}
\pi(\bm\mu) =& \int_{S(\delta^2)} \int_{S(\lambda)} \pi(\bm\mu \vert \lambda, \delta^2) \pi(\lambda) \pi(\delta^2) d\lambda d\delta^2 \ \ .
\end{align*}

\noindent
Consider only the following:
\begin{align*}
\pi(\bm{\mu}) =& \int_{0}^{\infty} \int_{-\infty}^{\infty} \pi(\bm{\mu} \vert \lambda, \delta^2) \pi(\lambda) \pi(\delta^2) d\lambda d\delta^2 \\ 
\propto& \int_{0}^{+\infty} \int_{-\infty}^{+\infty} (2\pi\delta^2)^{-\frac{m}{2}} \exp \bigg( -\frac{1}{2\delta^2} \sum_{i=1}^{m}(\mu_i-\lambda)^2 \bigg) \frac{1}{\delta^2} d\lambda d\delta^2 \\
\propto& \int_{0}^{+\infty} \bigg(\frac{1}{\delta^2}\bigg)^{\frac{m}{2}+1} \int_{-\infty}^{+\infty} \exp \bigg( -\frac{1}{2\delta^2} \sum_{i=1}^m(\mu_i^2-2\lambda\mu_i+\lambda^2) \bigg) d\lambda d\delta^2 \\
=& \int_{0}^{+\infty} \bigg(\frac{1}{\delta^2}\bigg)^{\frac{m}{2}+1} \int_{-\infty}^{+\infty} \exp \bigg( -\frac{1}{2\delta^2} \Big( \sum_{i=1}^{m}\mu_i^2 -2\lambda \sum_{i=1}^{m}\mu_i + m\lambda^2 \Big) \bigg) d\lambda d\delta^2 \ \ ;
\end{align*}
and set $\bar{\mu} = \frac{1}{m} \sum\limits_{i=1}^{m}\mu_i $; then, we obtain the following. 
\begin{align*}
\pi(\bm\mu) \propto& \int_{0}^{+\infty} \bigg(\frac{1}{\delta^2}\bigg)^{\frac{m}{2}+1} \exp \Big( -\frac{1}{2\delta^2}\sum_{i=1}^{m}\mu_i^2 \Big) \int_{-\infty}^{+\infty} \exp \bigg( -\frac{1}{2\frac{\delta^2}{m}} \Big(\lambda^2-2\lambda\bar{\mu} +\bar{\mu}^2 -\bar{\mu}^2 \Big) \bigg) d\lambda d\delta^2 = \\
=& \int_{0}^{+\infty} \bigg(\frac{1}{\delta^2}\bigg)^{\frac{m}{2}+1} \exp \bigg( -\frac{1}{2\delta^2} \Big( \sum_{i=1}^{m}\mu_i^2 -m\bar{\mu}^2 \Big) \bigg) \int_{-\infty}^{+\infty} \exp \bigg(-\frac{1}{2\frac{\delta^2}{m}} (\lambda-\bar{\mu}^2) \bigg) d\lambda d\delta^2 = \\
=& \int_{0}^{+\infty} \bigg(\frac{1}{\delta^2}\bigg)^{\frac{m}{2}+1} \exp \bigg( -\frac{1}{2\delta^2} m\Big( \frac{1}{m}\sum_{i=1}^{m}\mu_i^2 -\bar{\mu}^2 \Big) \bigg) \sqrt{2\pi\frac{\delta^2}{m}} d\delta^2 \propto \\
\propto& \int_{0}^{+\infty} \bigg(\frac{1}{\delta^2}\bigg)^{\frac{m-1}{2}+1} \exp \bigg(-\frac{1}{2\delta^2} \sum_{i=1}^{m}(\mu_i-\bar{\mu})^2 \bigg) d\delta^2 \ \ ,
\end{align*}

For any  choice of $m > 1$,  $\pi(\bm{\mu})$ can be written as follows. 
\begin{align*}
\pi(\bm\mu) \propto& \Big(\frac{1}{2} \sum_{i=1}^{m}(\mu_i-\bar{\mu})^2 \Big)^{-\frac{m-1}{2}} \Gamma \Big( \frac{m-1}{2} \Big) \propto 
\Big( \sum_{i=1}^{m}(\mu_i-\bar{\mu})^2 \Big)^{-\frac{m-1}{2}} \ \ .
\end{align*}

Now, we compute the following.
\begin{align*}
\pi(\bm{\gamma}) =& \int_{S(\sigma_1)}^{} \dots \int_{S(\sigma_m)}^{} \int_{S(\mu_1)}^{} \dots \int_{S(\mu_m)}^{} \pi(\bm{\gamma} \vert \bm\mu, \bm{\sigma^2}) \pi(\bm\mu) \pi(\bm{\sigma^2}) d\bm{\sigma}^2 d\bm\mu  \\
\propto& \int_{S(\sigma_1)}^{} \dots \int_{S(\sigma_m)}^{} \int_{S(\mu_1)}^{} \dots \int_{S(\mu_m)}^{}\prod_{i=1}^{m}\bigg[ (2\pi\sigma^2_i)^{-\frac{1}{2}} \exp \Big(-\frac{1}{2\sigma^2_i} (\gamma_i - \mu_i)^2 \Big) \bigg] \times  \\
& \frac{\prod_{i=1}^{m} (\sigma_i)^{-2}}{\Big( \sum_{i=1}^{m}(\mu_i-\bar{\mu})^2 \Big)^{\frac{m-1}{2}}} \,d\bm{\sigma}^2 d\bm\mu \\
\propto& \int_{S(\mu_1)}^{} \dots \int_{S(\mu_m)}^{} \Big( \sum_{i=1}^{m}(\mu_i-\bar{\mu})^2 \Big)^{-\frac{m-1}{2}} \prod_{i=1}^{m} \Bigg[ \int_{S(\sigma^2_i)}^{} \Big(\frac{1}{\sigma^2_i}\Big)^{\frac{3}{2}} \exp \Big( -\frac{1}{\sigma^2_i} \frac{(\gamma_i-\mu_i)^2}{2} \Big) d\sigma^2_i \Bigg] d\bm\mu  \\
\propto& \int_{S(\mu_1)}^{} \dots \int_{S(\mu_m)}^{} \Big( \sum_{i=1}^{m}(\mu_i-\bar{\mu})^2 \Big)^{-\frac{m-1}{2}} \prod_{i=1}^{m} \Big( (\gamma_i-\mu_i)^2 \Big)^{-\frac{1}{2}} d\bm\mu \\
=& \int_{S(\mu_1)}^{} \frac{1}{\vert\gamma_1-\mu_1\vert} \int_{S(\mu_2)}^{} \frac{1}{\vert\gamma_2-\mu_2\vert} \dots \int_{S(\mu_m)}^{} \frac{1}{\vert\gamma_m-\mu_m\vert}\frac{1}{\Big( \sum_{i=1}^{m}(\mu_i-\bar{\mu})^2 \Big)^{\frac{m-1}{2}}}  d\bm\mu .
\end{align*}

Notice that the following is the case:
\begin{align*}
\sum_{i=1}^{m} (\mu_i-\bar{\mu})^2 =& 
\sum_{i=1}^{m} \mu_i^2 - m \bar{\mu}^2 \\
=& \mu_m^2 + \sum_{i=1}^{m-1}\mu_i^2 - \frac{1}{m} \Big(\sum_{i=1}^{m}\mu_i\Big)^2 \\ 
=& \mu_m^2 + \sum_{i=1}^{m-1}\mu_i^2 - \frac{1}{m} \bigg( \Big(\sum_{i=1}^{m-1}\mu_i\Big)^2 + 2 \mu_m \Big( \sum_{i=1}^{m-1}\mu_i \Big) + \mu_m^2 \bigg);
\end{align*}
and set $K = \sum\limits_{i=1}^{m-1} \mu_i^2$ and $H = \sum\limits_{i=1}^{m-1} \mu_i$: then, we obtain the following. 
\begin{align*}
\sum_{i=1}^{m} (\mu_i-\bar{\mu})^2 =& 
\mu_m^2 + K - \frac{1}{m}(H^2 + 2H\mu_m+\mu_m^2) \\
=& \frac{m-1}{m}\mu_m^2 - \frac{2H}{m}\mu_m + K -\frac{1}{m}H^2 \ \ .
\end{align*}

\noindent
So $\sum\limits_{i=1}^{m} (\mu_i-\bar{\mu})^2$ is a convex parabolic function of $\mu_m$, and by the Weierstrass theorem, a global maximum exists for all bounded and closed sets.
By integrating  $\mu_m$, one obtains the following.

\begin{align*}
&\int_{S(\mu_m)}^{} \frac{1}{\vert\gamma_m-\mu_m\vert} \frac{1}{\Big( \sum_{i=1}^{m}(\mu_i-\bar{\mu})^2 \Big)^{\frac{m-1}{2}}}d\mu_m  \\
=& \int_{-\infty}^{+\infty} \frac{1}{\vert\gamma_m-\mu_m\vert}\frac{1}{\Big( \frac{m-1}{m}\mu_m^2 - \frac{2H}{m}\mu_m + K -\frac{1}{m}H^2 \Big)^{\frac{m-1}{2}}} d\mu_m \\
=& \int_{-\infty}^{\gamma_m} \frac{1}{\vert\gamma_m-\mu_m\vert}\frac{1}{\Big( \frac{m-1}{m}\mu_m^2 - \frac{2H}{m}\mu_m + K -\frac{1}{m}H^2 \Big)^{\frac{m-1}{2}}} d\mu_m \\
&+ \int_{\gamma_m}^{\epsilon} \frac{1}{\vert\gamma_m-\mu_m\vert}\frac{1}{\Big( \frac{m-1}{m}\mu_m^2 - \frac{2H}{m}\mu_m + K -\frac{1}{m}H^2 \Big)^{\frac{m-1}{2}}} d\mu_m  \\
&+ \int_{\epsilon}^{+\infty} \frac{1}{\vert\gamma_m-\mu_m\vert}\frac{1}{\Big( \frac{m-1}{m}\mu_m^2 - \frac{2H}{m}\mu_m + K -\frac{1}{m}H^2 \Big)^{\frac{m-1}{2}}} d\mu_m \ \ .
\end{align*}

\noindent
Let $A = \underset{\mu_m \in [\gamma_m, \epsilon]}{max}\Big( \frac{m-1}{m}\mu_m^2 - \frac{2H}{m}\mu_m + K -\frac{1}{m}H^2 \Big)$.
The second term of the last expression is as follows:
\begin{align*}
&\int_{\gamma_m}^{\epsilon} \frac{1}{\vert\gamma_m-\mu_m\vert}\frac{1}{\Big( \frac{m-1}{m}\mu_m^2 - \frac{2H}{m}\mu_m + K -\frac{1}{m}H^2 \Big)^{\frac{m-1}{2}}} d\mu_m \\
\ge& \int_{\gamma_m}^{\epsilon} \frac{1}{\vert\gamma_m-\mu_m\vert}\frac{1}{A^{\frac{m-1}{2}}} d\mu_m \\
=& \frac{1}{A^{\frac{m-1}{2}}} \int_{\gamma_m}^{\epsilon} \frac{1}{\mu_m-\gamma_m} d\mu_m  \\
=& \frac{1}{A^{\frac{m-1}{2}}} \Big[ \log(\mu_m-\gamma_m) \Big] \Big \vert_{\gamma_m}^{\epsilon} = +\infty \ \ ,
\end{align*}
which also implies the following. 
\begin{equation*}
\int_{S(\mu_m)}^{} \frac{1}{|\gamma_m-\mu_m|}\frac{1}{\Big[ \sum_{i=1}^{m}(\mu_i-\bar{\mu})^2 \Big]^{\frac{m-1}{2}}} d\mu_m = +\infty .
\end{equation*}

\noindent
For the same argument, one can also see that the following obtains.
\begin{equation*}
\pi(\bm\gamma)
\propto \int_{S(\mu_1)}^{} \frac 1{\vert\gamma_1-\mu_1\vert} \int_{S(\mu_2)}^{} \frac 1{\vert\gamma_2-\mu_2\vert} \dots \int_{S(\mu_m)}^{}\frac {\vert\gamma_m-\mu_m\vert^{-1}}{\Big[ \sum_{i=1}^{m}(\mu_i-\bar{\mu})^2 \Big]^{\frac{m-1}{2}}} 
d\bm\mu= +\infty \ \ ,
\end{equation*}

\noindent
It follows  that 
\begin{equation*}
\pi(\bm\gamma \vert U) \propto \prod_{i=1}^{m} \Big[ \prod_{j=1}^{n_i} \Big( c_i(U_{1ij}, U_{2ij}, \dots U_{d_iij} \vert \gamma_i)  \Big) \Big] \pi(\bm\gamma) = +\infty \ \ .
\end{equation*}

\noindent
A similar argument can be used to prove the following result.

{\bf Proposition 2:}
 If $\pi_{\sigma^2_i}(\sigma_i^2) \propto 1,$ for $i \in \{1, \dots, m\}$, and $\pi_{\delta^2}(\delta^2) \propto 1, \ \, \pi_{\lambda}(\lambda) \propto 1$, the posterior distribution $\bm\gamma \vert U$ is improper for any choice of copula densities   $c_i(\cdot \vert \gamma_i )$ and is independent of the sample size.

\noindent
{\bf Proof}:
As before,  one needs to show that the following pseudo-marginal posterior distribution of $\bm{\gamma}$ does  not have a finite integral.
\begin{align*}
\pi(\bm\gamma \vert U) =&
\int_{S(\bm\mu)} \int_{S(\bm{\sigma^2})} \int_{S(\delta^2)} \int_{S(\lambda)} \pi(\bm\gamma, \bm\mu, \bm{\sigma^2}, \lambda, \delta^2 \vert U) d\lambda d\delta^2 d\bm\sigma^2 
d\bm\mu \\
\propto& \prod_{i=1}^{m} 
\Big[  \prod_{j=1}^{n_i} \Big(  c_i(U_{1ij}, U_{2ij}, \dots , U_{d_i ij} \vert \gamma_i) \Big) \Big] \pi(\bm\gamma) \ \
\end{align*}

\noindent
We use the same notation as in Proposition 1 and assume $m > 3$ (when
$m\leq 3$, the theorem is trivially true since $\pi(\bm\mu)$ itself is not defined). With a slight modification in the proof of the proposition,  we obtain the following:
\begin{equation*}
\pi(\bm\mu) = \int_{S(\delta^2)}^{} \int_{S(\lambda)}^{} \pi(\bm\mu \vert \lambda, \delta^2) \pi(\lambda) \pi(\delta^2) d\lambda d\delta^2 \propto \Big( \sum_{i=1}^{m}(\mu_i-\bar{\mu})^2 \Big)^{-\frac{m-3}{2}} \ \ , 
\end{equation*}
and   \vspace{-6pt}
\begin{align*}
\pi(\bm\gamma) =& \int_{S(\sigma^2_1)}^{} \dots \int_{S(\sigma^2_m)}^{} \int_{S(\mu_1)}^{} \dots \int_{S(\mu_m)}^{} \pi(\bm\gamma \vert \bm\mu, \bm\sigma^2) \pi(\bm\mu) \pi(\bm\sigma^2) d\bm\mu d\bm\sigma^2\\
\propto& \int_{S(\sigma^2_1)}^{} \dots \int_{S(\sigma^2_m)}^{} \int_{S(\mu_1)}^{} \dots \int_{S(\mu_m)}^{} \prod_{i=1}^{m}\bigg[ (2\pi\sigma^2_i)^{-\frac{1}{2}} \exp \Big(-\frac{1}{2\sigma^2_i} (\gamma_i - \mu_i)^2 \Big) \bigg] \\
& \frac{1}{\Big( \sum_{i=1}^{m}(\mu_i-\bar{\mu})^2 \Big)^{\frac{m-3}{2}}} \, d\bm\mu d\bm\sigma^2\\ 
\propto& \int_{S(\mu_1)}^{} \dots \int_{S(\mu_m)}^{} \Big( \sum_{i=1}^{m}(\mu_i-\bar{\mu})^2 \Big)^{-\frac{m-3}{2}} \prod_{i=1}^{m} \Bigg[ \int_{S(\sigma^2_i)}^{} \Big(\frac{1}{\sigma^2_i}\Big)^{\frac{1}{2}} \exp \Big( -\frac{1}{\sigma^2_i} \frac{(\gamma_i-\mu_i)^2}{2} \Big) d\sigma^2_i \Bigg] d\bm\mu \ \
\end{align*}

\noindent
However, for all $i \in \{1, \dots, m\}$, the integral with respect to $\sigma^2_i$ is not finite, and this  again implies the following.
\begin{equation*}
\pi(\bm\gamma \vert U) \propto \prod_{i=1}^{m} \Big[ \prod_{j=1}^{n_i} \Big( c_i(U_{1ij}, U_{2ij}, \dots U_{d_iij} \vert \gamma_i)  \Big) \Big] \pi(\bm\gamma) = +\infty \ \ .
\end{equation*}
\end{appendices}

\end{document}